\documentstyle[aas2pp4]{article}
\newcommand{\ltwid}{\mathrel{\raise.3ex\hbox{$<$\kern-.75em\lower1ex\hbox{$\sim$}}}}
\newcommand{\gtwid}{\mathrel{\raise.3ex\hbox{$>$\kern-.75em\lower1ex\hbox{$\sim$}}}}
\newcommand{\s}{\scriptscriptstyle}
\newcommand{\sunn}{\s\odot}
\begin{document}
\title{Newly discovered Brown Dwarfs Not Seen in
Microlensing Time Scale Frequency Distribution?}
\author{S.J. Peale}
\affil{Dept of Physics\\
University of California\\
Santa Barbara, CA 93106\\
peale@io.physics.ucsb.edu} 
\begin{abstract}
The 2-Micron All Sky Survey (2MASS) (Skrutskie {\it et al.} 1997) and
the DEep Near Infrared Survey of the southern sky (DENIS) (Epchtein
{\it et al.} 1994) have revealed a heretofore unknown
population of free brown dwarfs that has extended the local mass
function down to as small as $0.01{\rm M}_{\sunn}$ (Reid {\it et al.}
1999).  If this local proportion of brown dwarfs extends throughout
the Galaxy---in particular in the Galactic bulge---one expects an
increase in the predicted fraction of short time scale microlensing
events in directions toward the Galactic bulge.  Zhao {\it et al.} (1996)
have indicated that a mass function with 30-60\% of the lens mass in
brown dwarfs is not consistent with empirical microlensing data. Here
we show that even the much lower mass fraction ($\sim 10\%$) of brown dwarfs
inferred from the new discoveries appears inconsistent with the
data. The added brown dwarfs do indeed increase the expected number of
short time scale events, but they appear to drive the peak in the time
scale frequency distribution to time scales smaller than that observed,
and do not otherwise match the observed distribution.  A reasonably
good match to the empirical data (Alcock {\it et al.} 1996) is
obtained by increasing the fraction of stars in the range
$0.08<m<0.7{\rm M}_{\sunn}$ considerably above that deduced from several
star counts. However, all inferences from microlensing about the
appropriate stellar mass function must be qualified by the meagerness
of the microlensing data and the uncertainties in the Galactic model. 
\end{abstract}
\subjectheadings{Galaxy: stellar content---stars: low-mass, brown dwarfs}
\section{Definitions and Assumptions}
A microlensing event is said to occur whenever a source star and lens
star pass each other at an angular separation that places the source
within an Einstein ring radius of the lens. The gravitational focusing
of the source light by the lens leads to the observed amplification of
the source brightness. The time scale of the event is defined as
$t_{\s E}=R_{\s E}/v$, where $R_{\s E}=\sqrt{4GmD_{\s OL}(D_{\s OS}-
D_{\s OL})/(c^2D_{\s OS})}$ is the Einstein Ring radius with $D_{\s
OL}$ and $D_{\s OS}$ being observer-lens and observer-source distances
respectively, $m$ the mass of the lens, $G$ the gravitational constant,
c the velocity of light, and $v$ is the relative transverse velocity
between lens and source projected onto the lens plane. Empirically, the
time scale is approximately half of the duration of an event.

To illustrate the effect of a population of brown dwarfs on the
microlensing time scale frequency distribution, we adopt the model of Zhao
(1996) for the mass distribution in a bar-like Galactic bulge,
\begin{equation}
\rho_{\s b}=\rho_{\s b0}\left[\exp\left(-\frac{s_{\s b}^2}{2}\right)+s_{\s
a}^{-1.85}\exp(-s_{\s a})\right], \label{eq:bulge}
\end{equation}
where 
\begin{eqnarray}
s_{\s b}^4&=&\left[\left(\frac{x}{\sigma_{\s x}}\right)^2+\left(\frac{y}
{\sigma_{\s y}}\right)^2\right]^2+\left(\frac{z^{\s\prime}}{\sigma_{\s
z}}\right)^4,\cr
s_{\s a}^2&=&\frac{q_{\s a}^2(x^2+y^2)+z^{{\s\prime}2}}{\sigma_{\s z}^2}
\label{eq:blgexp} 
\end{eqnarray}
with $q_{\s a}=0.6$, $\sigma_{\s x}=1.49$ kpc, $\sigma_{\s y}=0.58$ kpc and
$\sigma_{\s z}=0.40$ kpc. Here $x,y,z^{\s\prime}$ are orthogonal
coordinates along the principal axes of the triaxial bar.  The
coefficient $\rho_{\s b0}$ is chosen such that the mass of the bulge is
$2.2\times 10^{10}{\rm M}_{\sunn}$ (Zhao 1996).  To this we add the
exponential disk model of Bahcall \& Soneira (1980)  
\begin{equation}
\rho_{\s d}=\int_{m_{\s min}}^{m_{\s max}}
\frac{d\rho}{dm}dm=\rho_{\s
d0}\exp{\left(\frac{-|z^{\s\prime}|}{300{\rm pc}}-\frac{r}{s_{\s 
d}}\right)},   \label{eq:disk}
\end{equation}
where $z^{\s\prime}$ is the coordinate perpendicular to the plane of
the Galaxy and r is the radial coordinate in the plane of the Galaxy. We
choose $s_{\s d}=2.7$ kpc for the scale length in the radial direction
(Zhao, Spergel \& Rich 1995; Kent, Dame \& Fasio 1995) instead of 3.5
kpc chosen by Bahcall \& Soneira. The 
coefficient $\rho_{\s d0}$ is chosen such that $\rho_{\s 
d}=0.05{\rm M}_{\sunn}/{\rm pc^3}$ at $z^{\s\prime}=0$ and $r=8$ kpc,
the Sun's distance from the Galactic center.  The observations are
consistent with the long axis of the bulge inclined about $13^\circ$
to $20^\circ$ relative to the line of sight to the Galactic center
with the near side of the bar lying in the first quadrant (Zhao
1998). An inclination of $13^\circ$ is chosen in the following.

We normalize all distances by $D_8=8$ kpc, all velocity components
by $v_{\s\rm LSR}=210$ km/sec, the circular velocity of the local
standard of rest (LSR), the masses by the solar mass ${\rm M}_{\sunn}$
and the timescale $t_{\s E}$ by $t_0=\sqrt{4GM_{\sunn}D_8/(v_{\s\rm
LSR}^2c^2)}=66.72$ days, which is the timescale for an event with a
solar mass lens located at a distance of 4 kpc with a source at rest
at the Galactic center (8kpc).  For $t_0$, the lens velocity and observer
velocity are both assumed to be the circular velocity $v_{\s\rm
LSR}=210$ km/sec in the azimuthal direction consistent with a flat
Galaxy rotation curve. With these normalizations, the number of events
${\rm s^{-1}\;source^{-1}\;(unit}\,t_{\s E})^{-1}$ at timescale $t_E$
is (Peale 1998) 
\begin{eqnarray}
F&=&\frac{16\times10^{33}}{(3.0856\times
10^{18})^3}\sqrt{\frac{GD_8^3v_{\s LSR}^2}{M_{\sunn} c^2}}\int_{0.1}^{1.2} 
\int_0^\zeta\int_{-\infty}^\infty\times\cr
&& \int_{-\infty}^\infty \int_{-\infty}^\infty
\int_{-\infty}^\infty(v_{\s b}^2+v_{\s\ell}^2)\frac{d\rho_{\s L}(z,m)}{dm}
f_{\s v_{\s L\ell}^{\s\prime}}f_{\s v_{\s
Lb}^{\s\prime}}f_{\s v_{\s S\ell}^{\s\prime}}\times\cr
&&f_{\s v_{\s Sb}^{\s\prime}}\zeta^{2+2\beta}n_{\s S}(\zeta)
d\zeta\,dz\,dv_{\s L\ell}^{\s\prime}dv_{\s Lb}^{\s\prime}dv_{\s
S\ell}^{\s\prime}dv_{\s Sb}^{\s\prime}{\mbox{\Huge /}}\cr
&&\int_{0.1}^{1.2}\zeta^{2+2\beta}n_{\s S}(\zeta)d\zeta, \label{eq:mcarlo}
\end{eqnarray}
where $t_{\s E}$ appears only in the expression for $m$ in $d\rho_{\s
L} (z,m)/dm$ with
\begin{equation}
m=\frac{t_{\s E}^2\zeta(v_{\s b}^2+v_{\s\ell}^2)}{z(\zeta-z)}.\label{eq:m}
\end{equation} 
In Eqs. (\ref{eq:mcarlo}) and (\ref{eq:m}) $v_{\s
b},\,v_{\s\ell}$ are the components of the relative transverse
velocity $v$ in the directions of increasing Galactic latitude and
longitude respectively, $\rho_{\s L}(z,m)$ is the lens mass density, where the
numerical coefficients in Eq. (\ref{eq:mcarlo}) mean it is expressed
in ${\rm M}_{\sunn}/{\rm pc^3}$, $\zeta=D_{\s OS}/D_{\s 8}$, $z=D_{\s
OL}/D_{\s 8}$. The distributions of random velocities about the
circular velocity of the Galactic model are $f_{v^{\s\prime}_{\s Xy}}=
\exp{(-v^{{\s\prime}2}_{\s Xy}/(2\sigma_{\s
Xy}^2))}/(\sqrt{2\pi}\sigma_{\s Xy})$, where $\sigma_{\s Xy}$ is the rms
value of the random velocity $v^{\s\prime}_{\s Xy}$ and the subscript
$X$ is either $L$ or $S$ for lens or source and the subscript
$y$ is $b$ or $\ell$ for the latitude and longitude directions. The
spatial density of sources is $n_{\s S}(\zeta)$, and the exponent
$2+2\beta$ follows from the increase in area of the cross section of
the angular field with distance and from the assumption that the
fraction of stars with luminosities greater than some $L_*$ varies as
$L_*^{\s\beta}$ (Kiraga \& Paczy\'nski 1994).  

The details of the
derivation of Eq. (\ref{eq:mcarlo}) are given in Peale (1998). 
The expression for the rate of events for a single source at $D_{\s
OS}$ that has random velocity components within $dv^{\s\prime}_{\s
Sb}dv^{\s\prime}_{\s S\ell}$ of ($v_{\s Sb}^{\s\prime},v_{\s
S\ell}^{\s\prime}$) for lenses within $dD_{\s OL}$ of $D_{\s
OL}$ with random velocities within $dv^{\s\prime}_{\s
Lb}dv^{\s\prime}_{\s L\ell}$ of ($v_{\s Lb}^{\s\prime},v_{\s
L\ell}^{\s\prime}$), and with lens mass within $dm$ of $m$ is
multiplied by the fraction of sources within $dD_{\s OS}$ of $D_{\s OS}$
and by $\delta(t_{\s E}-t_{\s E}^{\s\prime})$ to select a
particular time scale. Independent variable $m$ is replaced by $t_{\s
E}$ and the expression integrated over the ranges of all the variables
as a function of $t_{\s E}^{\s\prime}$ to yield the time scale frequency
distribution of Eq. (\ref{eq:mcarlo}). The integral limits correspond
to the variables in the same order as the differentials in the
integrand, where the integration over $\zeta$ starts at 0.1 instead of
zero to avoid a singularity.  This is justified as there are not
likely to be any sources involved in microlensing events closer to us
than 0.8 kpc. The distributions of lenses and sources along a
particular line of sight are identical and follow from the above
Galactic model.  The assumed circular velocities for the general
Galactic rotation are $v_{\s L0},v_{\s S0} = 210$ km/sec for
$z,\zeta<0.5875$ ($r>3.3$ kpc) consistent with a flat rotation curve
except the circular velocities decrease linearly to 0 at the Galactic
center for $0.5875<z,\zeta<1.0$ ($r<3.3$ kpc).  The velocity
dispersions in the disk are $(\sigma_{\s db},\sigma_{\s
d\ell})=(16,20)$ km/sec ($z,\zeta<0.5875$), whereas the bulge velocity
dispersions are $(\sigma_{\s bb},\sigma_{\s b\ell})=(110,110)$ km/sec
($0.5875<z,\zeta<1.0$).  The velocity dispersions $\sigma_{\s
Lb},\sigma_{\s L\ell},\sigma_{\s Sb},\sigma_{\s S\ell}$ are assigned
to disk or bulge dispersions in Eq. (\ref{eq:mcarlo})  according to
the locations of the lens and source.  Evaluation of the integral
with a Monte Carlo technique is described in Peale (1998).  
\section{Results}
Estimating the mass function for brown dwarfs is a complicated
process, since, unlike main sequence stars, there is no unique
correspondence between mass and spectral type.  The spectral type
L dwarfs, observations of which are the basis of the Reid {\it et al.}
(1999) discussion, have effective temperatures between 1400 and 2000 K
with no $\rm CH_4$ absorption, but every brown dwarf with
$0.01<m<0.075M_{\s\odot}$ passes through the L dwarf region as it
cools, with those of higher mass passing through later (Burrows, {\it
et al.} 1997). The number of
L-type brown dwarfs at any epoch depends on the birth rate, the
initial mass function, and the evolution of these failed
stars (Reid, {\it et al.} 1999). If the birth rate is assumed known
and a theory of evolution adopted, the dependence of the number of
visible L dwarfs on the mass function can be used to constrain the
latter.  If a power law distribution of mass is assumed, the steeper
distributions lead to more visible L dwarfs, and the index is
constrained by the observed volume density of L dwarfs in the region
reasonably close ($<8$ pc) to the Sun.  This procedure leads to
$dn/dm\sim m^{-1.3}$ for $0.01<m<0.08{\rm M}_{\sunn}$ based on
information from the DENIS and 2MASS surveys (Reid {\it et al.} 1999).
Although model dependent, this result is the first determination of
the brown dwarf mass function based on observational data. What are
the implications for the microlensing time scale frequency
distribution?

Several star counts yield $dn/dm\sim m^{-1}$ for 
$0.08<m<\sim 0.7-1.0{\rm M_{\sunn}}$ with steeper slopes for more
massive stars (e.g. Basu \& Rana 1992; Holtzman {\it et al.} 1998;
Reid {\it et al.} 1999).  We shall use the Holtzman {\it et al.} mass
functions and modifications thereof in the examples. Figure 1 shows
the time scale frequency distributions for the above Galactic model
for a line of sight toward $(\ell,b)=(1^\circ,-4^\circ)$, which is
representative of an average (see Peale 1998) over the directions to
24 fields observed by the MACHO (MAssive Compact Halo Objects) group
during the 1993 bulge season. During this 
season 12.6 million  stars were monitored for 189 days (Alcock {\it et
al.} 1996) leading to the choice of ordinate in Figure 1. The 
distributions are shown for the Holtzman {\it et al.} (1998) mass
function (solid line) and three modifications, two of which include
brown dwarfs down to $\rm 0.01M_{\sunn}$ to accommodate the new
discoveries. For continuation of the main sequence mass function into the
brown dwarf region with slope -1, the mass of the brown dwarfs is only
5.5\% of that of the main sequence stars, and it is about 7\% of the
main sequence star mass with the Reid {\it et al.} slope of -1.3, where
continuity of the mass function is assumed.  These brown dwarf masses
are 8\% and 10\%, respectively, of those main sequence stars with
$0.08\le m\le 1.0M_{\sunn}$---a range more appropriate to the solar
neighborhood.  The third modification contains no brown dwarfs but has
many more low  
mass, hydrogen burning stars, and it yields a time scale frequency
distribution  closely matching the empirical data. (The peaks of these
curves are raised by  about 10\% if the value of $v_{\s LSR}=270$
km/sec found by Mendez {\it et al.} (1999) is appropriate for the flat
part of the Galaxy rotation curve.) All the mass 
functions are continuous and are normalized to yield a bulge mass of
$2.2\times 10^{10}{\rm M_{\sunn}}$ and a mass density at the Sun's
position in the disk of $0.05{\rm M_{\sunn}/pc^3}$. The empirical time
scale frequency distribution from the MACHO 1993 bulge season (Alcock
{\it et al.} 1996), adjusted for observing efficiencies and binned
into 5 day intervals, is shown for comparison. 
\begin{figure}[ht]
\plotone{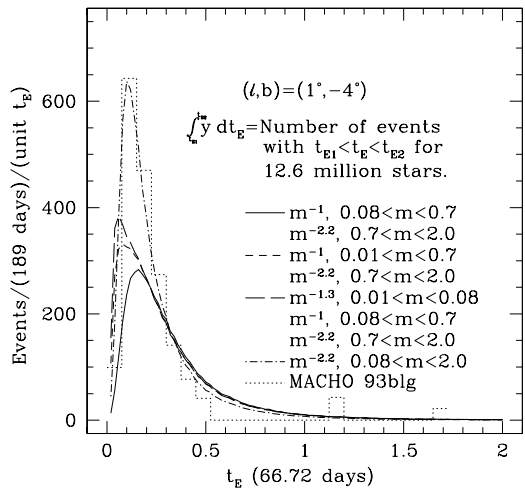}
\caption {Comparison of predicted microlensing time
scale frequency distributions with the MACHO observational data
corrected for observing efficiencies.  The Zhao (1996) density
distribution of the triaxial galactic bulge and the Bahcall-Soneira
(1980) density distribution of the disk (with disk scale length 2.7
kpc) are assumed. The average rotation curve of the galaxy is assumed
flat at 210 km/sec for galactocentric distance $r>3.3$ kpc and
declining linearly to zero at the galactic center for $r<3.3$
kpc. Velocity dispersions in the directions of increasing galactic
latitude and longitude about the circular motion are $(\sigma_{\s
db},\sigma_{\s d\ell})=(16,20)$ km/sec for $r>3.3$ kpc and
$(\sigma_{\s bb},\sigma_{\s b\ell})=(110,110)$ km/sec for $r<3.3$ kpc.}
\end{figure}

As found by several authors (e.g. Peale 1998), mass functions
with $dn/dm\sim m^{-1}$ in the M star
region fail to yield a sufficient number of short time
scale events near the peak of the empirical distribution {\it for the
simplest Galactic models consistent with observations}. Adding the
Reid {\it et al.} brown dwarfs to the Holtzman {\it  et al.}
distribution with either the -1.3 or -1 slope increases the 
number of short time scale events as expected, but both extensions of
the mass function fail to yield a sufficient number of short time
scale events near the peak of the Alcock {\it et al.} results, and,
perhaps more importantly, the peaks of the predicted distributions are
shifted toward smaller time scales (3-5 days) than those of the
observed peak (5-10 days). (See Peale (1998) for similar consequences
for other extensions of the mass function into the brown
dwarf region.)  A fairly good match to the
empirical data is obtained in Figure 1 by increasing the fraction of
low mass, but still hydrogen burning stars with a continuation of the
-2.2 slope of the Holtzman {\it et al.} mass function down to
$m=0.08{\rm M_{\sunn}}$. 
\section{Discussion}
The chosen model of the Galaxy used in calculating the predicted time scale
frequency distributions is consistent with observations
(Zhao 1996, 1998), although many parameters are poorly constrained
(Zhao 1998).  If the Galaxy model is assumed to be close to the real
Galaxy, and the local population of brown dwarfs extends throughout the
Galaxy, then the MACHO collaboration should have seen a larger
fraction of events with 3-5 day time scales.  One could conclude that
the absence of these events means that there are fewer brown dwarfs in
the bulge, where most of the lenses reside, than observed
locally. However, the same caveats and reservations about microlensing
constraints on the mass function emphasized by Peale (1998) apply
here.  

The meagerness of the published microlensing data for bulge
events precludes definitive microlensing constraints on the
distribution of stars in the Galaxy or the mass function, although
there are more than 200 events that have not yet been analyzed that
would improve this situation.  We also point out that the 10 events in
the bin corresponding to time scales from 5 to 10 days in Figure 1
involve two clump giant sources with detection efficiencies of about
0.6 and eight much less luminous turnoff and possibly more blended
stars as sources with detection efficiencies of 0.15 to 0.2 (Alcock
{\it et
al.} 1996).  The average of the reciprocals of the efficiencies is
nearly 5 meaning this element of the histogram is almost 5 times higher
than the actual number of detected events. The sampling efficiencies
for a given time scale event for a distribution of source 
brightnesses are determined empirically and should be reasonably
sound, but the reduction of these efficiencies due to blending of the 
source stars with other nearby stars by a factor of 0.75 is just an
estimate (Alcock et al 1996).  Until the detection efficiencies for
bulge events are determined more rigorously, even the empirical
distribution of MACHO time scales that we are trying to match is only
a first approximation.  Could the detection efficiency for the 3-5 day
events be even smaller than the Alcock {\it et al.} estimate?  If so, this
element of the histogram would be raised, although it seems
unlikely that this region could become the peak of the distribution
and thereby accommodate the brown dwarfs.

If the empirical distribution approximates reality, the brown 
dwarfs might be accommodated by reducing the velocity dispersion (See
examples in Peale 1998), but unrealistically low dispersions would be
necessary to shift the peak to the MACHO value, and increasing the
mass of the Galaxy to yield the observed peak frequency of events
would result in too many events with $t_{\s E}\gtwid 20$
days. Finally, one notices the several very long time scale events in
the MACHO distribution that are not consistent with any of the
models. These events may be due to there being more close lenses, such
as a spiral arm concentration, where the relative angular velocity of
source and lens is small, than are accounted for in the axisymmetric
disk model.  Other than the triaxial bulge, deviations from
axial symmetry in the Galaxy model are ignored. In addition, streaming
velocities superposed on the Galactic rotation are not accounted
for. Could some combination of such additional degrees of freedom hide
a Galaxy-wide distribution of brown dwarfs in the microlensing data?
One cannot be sure without better constraints on Galactic structure
and velocity parameters. For the present, the addition of the
brown dwarfs to the mass function appears to widen the discrepancies
between predicted and observed microlensing time scale frequency
distributions. 
\begin{center}{\bf Acknowledgements}\end{center}
Thanks are due Omer Blaes for suggesting additional content to the
manuscript. This work is supported by NASA OSS program under grant 
NAG5-7177. 

\begin{center}{\bf References}\end{center}
\begin{description}
\item Alcock, C., et al. 1996, ApJ,  479, 119

\item Bahcall, J. N. \& Soneira, R. M. 1980, ApJSup, 44, 73

\item Basu, S. \& Rana, N. C. 1992, ApJ, 393, 373

\item Burrows, A., Marley, M., Hubbard, W. B., Lunine, J.I. {\it et
al.} 1997. ApJ, 491, 856

\item Epchtein, N., De Batz, B., Copet, E. et al. 1994, in {\it Science with
Astronomical Near-Infred Sky Surveys}, ed. N. Epchtein, A. Omont,
B. Burton, P. Persei, (Kluwer, Dordrecht), p. 3

\item Holtzman, J. A., Watson, A. M., Baum, W. A., Grillmair, C. J.,
et al. 1998, AJ, 115, 1946 

\item Kent, S. M., Dame T. M. \& Fasio, G. 1991, ApJ, 378, 131

\item Kiraga, M. \& Paczy\'nski, B. 1994, ApJ, 430, L101

\item Peale, S. J. 1998, ApJ, 509, 177

\item Mendez, R. A., Girard, T. M., Kozhurina-Platais, V., van Altena,
W. F. et al. 1999, Submitted to Apj Lett.

\item Reid, I. N., Kirkpatrick, J.D., Liebert, J. Burrows, A. et al. 1999,
ApJ, 521, 613

\item Skrutskie, M. F., Schneider, S. E., Stiening, R., Strom, S. E.,
et al. 1997, in The Impact of Large-Scale Near-IR Sky Survey,
ed. F. Garzon et al. (Kluwer, Dordrecht), p. 25 

\item Zhao, H. S. 1996, MNRAS, 283, 149

\item Zhao, H. S. 1998, MNRAS, In Press

\item Zhao, H. S., Spergel, D. N., \& Rich, R. M. 1995, ApJ, 440, L13

\item Zhao, H. S., Rich, R. M., \& Spergel, D. N. 1996, MNRAS, 282, 175
\end{description}

\end{document}